\documentclass[usenatbib]{mn2e}
\usepackage{epsfig,lscape,color}

\newcommand{\etal}{et~al.} 

\newcommand{\UCHII}{UCH{\sc ii} }

\newcommand{\kms}{$\mbox{km~s}^{-1}$ }
\newcommand{\kmsns}{$\mbox{km~s}^{-1}$}

\newcommand{\Msol}{M$_{\odot}$}
\newcommand{\Lsol}{L$_{\odot}$}

\newcommand{\vsfig}[2]           
{
  \begin{center}
    \begin{minipage}[t]{0.05\textwidth}
      {\footnotesize \raisebox{40mm}{(#2)}}
    \end{minipage}
    \begin{minipage}[t]{0.42\textwidth}
      \psfig{file=./#1.ps,height=0.95\textwidth,angle=0}
    \end{minipage}
    \hfill
  \end{center}
}

\newcommand{\specdfig}[2]        
{
   \begin{center}
     \begin{minipage}[t]{0.45\textwidth}
         \psfig{file=eps/#1.eps,height=0.65\textwidth,width=1\textwidth,angle=0}
     \end{minipage}
     \hfill
     \begin{minipage}[t]{0.45\textwidth}
         \psfig{file=eps/#2.eps,height=0.65\textwidth,width=1\textwidth,angle=0}
     \end{minipage}
   \end{center}
}

\newcommand{\specsfig}[1]        
{
   \begin{center}
     \begin{minipage}[t]{0.45\textwidth}
         \psfig{file=eps/#1.eps,height=0.65\textwidth,width=1\textwidth,angle=0}
     \end{minipage}
   \end{center}
}

\newcommand{\boxfig}[1]        
{
   \begin{center}
     \begin{minipage}[t]{0.46\textwidth}
         \psfig{file=eps/#1.eps,height=0.45\textwidth,angle=0}
     \end{minipage}
   \end{center}
}

\newcommand{\twofig}[2]        
{
   \begin{center}
     \begin{minipage}[t]{0.5\textwidth}
         \psfig{file=eps/#1.eps,height=0.95\textwidth}
     \end{minipage}
     \hfill
     \begin{minipage}[t]{0.5\textwidth}
         \psfig{file=eps/#2.eps,height=0.95\textwidth}
     \end{minipage}
   \end{center}
}

\begin{document}

\title[New water masers in the SMC]{Discovery of four water masers in the Small Magellanic Cloud}
\author[S.\ L.\ Breen \etal]{S.\ L. Breen,$^{1}$\thanks{Email: Shari.Breen@csiro.au} J.\ E.\ J.\ Lovell,$^2$ S.\ P. Ellingsen,$^2$ S.\ Horiuchi,$^3$ A.\ J.\ Beasley,$^4$ \newauthor K.\ Marvel$^5$ \\
  \\
  $^1$ CSIRO Astronomy and Space Science, Australia Telescope National Facility, PO Box 76, Epping, NSW 1710, Australia\\
  $^2$ School of Mathematics and Physics, University of Tasmania, Private Bag 37, Hobart, Tasmania 7001, Australia\\
  $^3$ CSIRO Astronomy and Space Science, Canberra Deep Space Communications Complex, PO Box 1035, Tuggeranong, ACT 2901, Australia\\
  $^4$ National Radio Astronomy Observatory, 520 Edgemont Road Charlottesville, VA 22903-2475, USA \\
  $^5$ American Astronomical Society, 2000 Florida Avenue, NW, Suite 400, Washington, DC 20009, USA\\
}
 
 \maketitle
  
 \begin{abstract}
 We report the detection of four water masers within the Small Magellanic Cloud (SMC); two discovered with the 70-m Tidbinbilla radio telescope, and two discovered with the Australia Telescope Compact Array (ATCA). Precise positions of all four masers have been derived from ATCA observations, and the characteristics of each water maser have been monitored over a period of several years. Sensitive observations towards two previously detected water masers reported in the literature failed to detect any emission. The detected water masers show evidence of higher levels of temporal variability than equivalent Galactic sources, and one of the features associated with NGC346~IR1 shows an acceleration of  9.6~\kms~yr$^{-1}$ over a 31 day period. Sensitive targeted observations for methanol and OH masers failed to detect any accompanying emission - in the case of methanol perhaps highlighting an under abundance in the SMC, consistent with expectations due to lower metallicity. 

The water masers are both bright and compact making them excellent targets for Very Long Baseline Interferometry (VLBI) observations, which can potentially measure the proper motion of the SMC ($\sim$1--2~mas~yr$^{-1}$) with temporal baselines of $\sim$12 months.  Such observations would utilise sources associated with only the current epoch of star formation and hence have several advantages over alternative methods.

\end{abstract}


\section{Introduction}

The Small Magellanic Cloud (SMC) is a dwarf, irregular galaxy at a
distance of approximately 61 kpc.  It has a metallicity less than
one-fifth solar \citep{Peimbert00} and is considered an ideal
laboratory for studying star formation from primordial gas.  Although
the mass of the SMC is approximately 1/250th the mass of the Milky Way
\citep{Stan04}, its current star formation rate of 0.05
solar masses per year \citep{Wilke04} is approximately 1/100th
that of our Galaxy \citep{Diehl06}.  The reason for the enhanced
rate of star formation in the SMC at the current epoch (in comparison
to the LMC and Galaxy) is unknown, but likely to be related to
interaction of the Magellanic system with our Galaxy \citep[e.g.][]{Diaz12}.

One of the key tracers of young star formation regions in our own Galaxy are interstellar masers \citep[e.g.][]{Ellingsen06}. Within these regions, masers of water, methanol and hydroxyl are common and strong \citep[e.g.][]{Breen10,GreenMMB10,C98}, making them excellent tools for understanding properties of the associated star formation. Some maser species accurately trace the systemic velocities of the regions that they are associated with \citep[e.g.][]{Szy07,Pandian09,GM11}, increasing their application to studies of Galactic structure \citep[e.g.][]{Green11}, while others are thought to trace the evolution of high-mass star forming regions \citep[e.g.][]{Ellingsen07,Breen10b}.

To date only two sources with masers (of any species) have been observed in the SMC and these were discovered more
than 30 years ago in a search for water maser emission towards 15 regions with the Itapetinga radio telescope \citep{SB82}. Other extensive searches for masers have failed to detect any emission from the 6.7-GHz transition of methanol \citep[e.g.][]{Ellingsen94,Beasley96,Green08}, or OH maser emission \citep[e.g][]{Green08}. A full account of previous observations for both interstellar and circumstellar masers in the Magellanic Clouds has been reported in \citet{vanLoon12}.

Observations of masers within the SMC not only allow young high-mass star formation regions to be pinpointed, but can also enable the populations as a whole to be studied, and compared with the populations in other nearby galaxies (including our own), allowing the effects of environmental conditions on maser populations to be studied. \citet{Green08} used their complete methanol maser survey of the LMC and SMC to conclude that the methanol masers in the LMC were under abundant by a factor of 4-5 (once the varying star formation rates were accounted for), likely because of the lower metallictiy in the LMC. Whereas \citet{Brunth06} compared the number of detected water masers in the Local Group, finding that detection rates are consistent with those of our Galaxy, with the exception of IC 10 which exhibited an overabundance. Unlike the 6.7-GHz methanol maser samples used in \citet{Green08}, the water maser samples used in \citet{Brunth06} lack completeness, leaving a lot of uncertainties in both the number of sources expected and the numbers detected. In a modestly sized searches of star formation regions in the LMC, \citet{Ellingsen2010} discovered six new water masers, while \citet{Imai13} found a further eight, highlighting the incompleteness of the samples used in previous analyses.  

Arguably the greatest impact that the currently incomplete sample of masers in the Local Group can provide is through measurements of accurate distances and dynamics of its members through Very Long Baseline Interferometry (VLBI) observations. Such measurements are ongoing in our own Galaxy \citep[e.g.][]{Reid09,Honma12}, incorporating an increasingly large number of methanol masers spread throughout the Galactic plane. The increasing number of masers detected in other galaxies of the Local Group mean that it is now feasible to begin incorporating proper motion studies of additional members into such programs. Water masers are not favoured for such VLBI programs in our Galaxy due to their often extreme temporal variability \citep[e.g.][]{Brand03,Felli07} and their tendency to trace more complex internal source dynamics (such as outflows). However, when measurements are made using both water and methanol masers they can be consistent to within a few per cent \citep[e.g. W3(OH):][]{Reid09,Hach06} and their relatively higher detection rates, flux densities and often simple spectra make them desirable targets in other members of the Local Group. Through proper motion measurements, masers provide the best opportunity to understand the present state and therefore the history and future of the Local group.

Here we present a series of observations of water masers within the SMC. The primary motivation of these observations was to identify and characterise water masers in the SMC that can be targets for future astrometric VLBI programs to measure proper motions of the SMC in the short term, and possibly  to measure trigonometric parallax once astronometric accuracies have increased to make such measurements feasible. The water masers we detected have been accurately positioned with the Australia Telescope Compact Array (ATCA), and monitored to assess the feasibility of a long term VLBI program.

\section{Observations and data reduction}

Between 2003 and 2011 we conducted a number of water maser observations in the SMC; both searching for new water maser sites and monitoring the emission from detected sources. The water maser observations used unallocated Host-Country time on the Tidbinbilla 70 m radio telescope and Director's time on the ATCA. In general only short allocations of time were available so the observing strategy and number of sources observed was adapted for each observation. We supplemented our water maser observations with an additional epoch of ATCA data (project code C973) from the Australia Telescope Online Archive (ATOA). Table~\ref{tab:obs} lists the epochs each source was observed, the telescope used, the spectral resolution, RMS noise of the observations and the characteristics of detected emission. Targeted observations for OH and methanol masers were also made towards each of the detected water masers using the Parkes 64 m radio telescope and the ATCA (these are described in Section~\ref{sect:other}).

\begin{table*}
\begin{center}
  \caption{Summary of observations and characteristics of targeted water maser emission. Accurate positions for the four water masers we detect are listed in Table~\ref{tab:detections}. Column 1 gives the source name, column 2 gives the observing date, column 3 shows the telescope used (and array configuration if the ATCA), columns 4-6 give the 1-$\sigma$ sensitivity, spectral resolution and velocity coverage of the observations. Columns 7-9 give the characteristics of the detected water masers emission at each observation and includes the peak flux density (or 5-$\sigma$ detection limit if not detected), velocity (heliocentric) of the peak emission and the minimum and maximum velocity of the detected emission. }
  \begin{tabular}{lllcccccccc}\hline\label{tab:obs}
 {\bf Source} &{\bf Date} & {\bf Telescope}& {\bf 1-$\sigma$} & {\bf Spectral} & {\bf Velocity} & {\bf Speak} & {\bf Vpeak} & {\bf Vrange}\\
		 & {\bf (YYDOY)}  & {\bf \& array} & {\bf sensitivity}& {\bf resolution} &{\bf coverage}  & \\ 
		 &			&			&	{\bf (mJy)}	&	{\bf (\kmsns)} &	{\bf (\kmsns)}	&{\bf (Jy)}&	{\bf (\kmsns)} &	{\bf (\kmsns)} \\ \hline
NGC346 IR1		&	03123	&	ATCA,EW352	&	58	&	0.5	&	10,220	&	$<$0.29	&	--	&	--\\
				&	08200	&	Tidbinbilla	&	39	&	0.5	&	--470,600	&	2.6	&	173.3	&	168.2,174.2	\\
				&	08203	&	ATCA,H214	&	44	&	0.25	&	70,280	&	3.4	&	173.4	&	167.9,173.8	\\
				&	08203	&	Tidbinbilla	&	108	&	0.06	&	--10,250	&	2.7	&	173.3	&	167.9,173.9	\\
				&	08211	&	Tidbinbilla	&	159	&	0.06	&	--10,250	&	1.3	&	168.5	&	167.8,169.1	\\
				&	08213	&	ATCA,H214	&	24	&	0.25	&	70,280	&	1.2	&	168.4	&	167.8,169.1	\\
				& 	08215	&	Tidbinbilla	&204	&	0.06	&	0,250	&	1.4	&	168.1	&	167.6,169.0	\\
				&	08218	&	Tidbinbilla	&	149	&	0.06	&	0,250	&	1.3	&	168.0	&	167.6,168.8	\\
				&	08228	&	ATCA,6B		&	78	&	0.25	&	60,270	&	1.0	&	168.0	&	167.3,168.6	\\
				&	08229	&	ATCA,6B		&	26	&	0.25	&	60,270	&	0.25	&	168.0	&	167.3,168.6	\\
				&	08231	&	ATCA,6B		&	23	&	0.25	&	60,270	&	0.92	&	167.8	&	167.2,168.5	\\
				& 	08232	&	ATCA,6B		&	25	&	0.25	&	60,270	&	1.0	&	167.9	&	167.3,168.7	\\
				& 	11153-5	&	ATCA,H214	& 	31	&	0.5	&	--415,440	&	0.24	&	167.4	&	166.3,173.3	\\
\\
N66 Sc12			&	03113	&	Tidbinbilla	&	177	&	0.06	&	45,255	&	1.6	&	164.7	&	164.3,165.0	\\
				&	03123	&	ATCA,EW352	&	18	&	0.5	&	10,220	&	1.6	&	164.5	&	163.7,166.2	\\
				&	08203	&	ATCA,H214	&	35	&	0.25	&	70,280	&	0.46	&	163.9	&	163.5,174.2	\\
				&	08211	&	Tidbinbilla	&	70	&	0.06	&	--10,250	&	0.30	&	166.3	&	163.5,174.3	\\
				& 	08213	&	ATCA,H214	&	28	&	0.25	&	70,280	&	0.18	&	164.0	&	164.0,174.3	\\
				&	08215	&	Tidbinbilla	&	76	&	0.06	&	0,250	&	0.31	&	166.1	&	166.1,174.1	\\
				&	08218	&	Tidbinbilla	&	107	&	0.06	&	0,250	&	0.30	&	166.1	&	163.8,166.1	\\
				&	08228	&	ATCA,6B		&	29	&	0.25	&	60,270	&	0.29	&	174.2	&	163.5,174.3	\\
				&	08229	&	ATCA,6B		&	27	&	0.25	&	60,270	&	0.24	&	174.1	&	162.7,174.1	\\
				&	08231	&	ATCA,6B		&	18	&	0.25	&	60,270	&	0.29	&	166.2	&	163.4,174.4	\\
				& 	08232	&	ATCA,6B		&	20	&	0.25	&	60,270	&	0.28	&	166.2	&	163.5,174.4	\\
				& 	11153-5	&	ATCA,H214	&	31	&	0.5	&	--415,440	&	0.29	&	161.1	&	160.2,167.0	\\
				\\
IRAS00430--7326	& 	03123	&	ATCA,EW352	&	33	&	0.5	&	10,220	&	2.1	&	138.6	&	137.4,142.6	\\
				&	08213	&	ATCA,H214	&	32	&	0.25	&	70,280	&	2.4	&	141.6	&	136.9,142.6	\\
				&	08228	&	ATCA,6B		&	51	&	0.25	&	60,270	&	1.9	&	141.5	&	136.7,142.6	\\
				&	08229	&	ATCA,6B		&	26	&	0.25	&	60,270	&	2.2	&	141.5	&	137.1,142.8	\\
				&	08231	&	ATCA,6B		&	32	&	0.25	&	60,270	&	1.9	&	141.6	&	137.0,143.1	\\
				& 	08232	&	ATCA,6B		&	43	&	0.25	&	60,270	&	2.7	&	141.4	&	136.4,142.7	\\
				& 	11153-5	&	ATCA,H214	&	30		&	0.5	&	--415,440	&	1.8	&	141.8	&	137.6,142.4	\\ 
				\\
IRAS01126--7332	& 	03123	&	ATCA,EW352	&	32	&	0.5	&	10,220	&	1.7	&	177.4	&	176.0,178.4	\\
				&	08213	&	ATCA, H214	&	34	&	0.25	&	70,280	&	0.51	&	177.3	&	176.7,178.0	\\
				&	08215	&	Tidbinbilla	&	262	&	0.06	&	0,250	&	0.76	&	177.3	&	176.7,181.7	\\
				&	08228	&	ATCA,6B		&	78	&	0.25	&	60,270	&	0.80	&	177.5	&	176.6,181.9	\\
				&	08229	&	ATCA,6B		&	28	&	0.25	&	60,270	&	0.39	&	177.3	&	176.6,181.9	\\
				&	08231	&	ATCA,6B		&	25	&	0.25	&	60,270	&	0.71	&	177.2	&	176.4,181.8	\\
				& 	08232	&	ATCA,6B		&	23	&	0.25	&	60,270	&	0.76	&	177.3	&	176.4,181.9	\\
				& 	11153-5	&	ATCA,H214	&	30	&	0.5	&	--415,440	&	0.68	&	177.0	&	175.3,181.6	\\
				\\
s7				&	03123	&	ATCA,EW352	& 	35	&	0.5	&	10,220	&	$<$0.18 &		--	&	--\\
				&	08173	&	Tidbinbilla	&	24	&	0.5	& --550,700	&	$<$0.12	&		--	&	--\\
				&	08213	&	ATCA,H214	&	25 	&	0.25	&	70,280	&	$<$0.13	&	--	&	--	\\
				\\
s9				&	03123	&	ATCA,EW352	&	35	&	0.5	&	10,220	&	$<$0.18	&	--	&	--	\\
				&	08173	&	Tidbinbilla	&	26	&	0.5	&--550,700	&	$<$0.13	&	--	&	--	\\
				&	08213	&	ATCA,H214	&	25 	&	0.25	&	70,280	&	$<$0.13	&	--	&	--	\\\\
\hline
	\end{tabular}
\end{center}

\end{table*}

\subsection{Tidbinbilla observations}

The primary search for water masers in the SMC was undertaken in two phases; the first, a complete search of a 5 x 5 arcmin grid centred on 00$h$ 59$m$ 21.25$s$ --72$^{\circ}$ 11$'$ 03$"$ (marked on Fig.~\ref{fig:3col}), and the second, a targeted search towards a number of bright infrared sources in the NGC346 region \citep[][listed in Table~\ref{tab:bright24}]{Simon07}. Both searches were made using the Tidbinbilla 70-m radio telescope, which forms part of NASAs Deep Space Network, during unallocated Host-Country time. Observations were made using the K-Band receiver that detects both left and right circularly polarised signals. Pointing errors were measured and corrected by scanning bright quasars at the nearest positions in the sky from the targets, yielding residual pointing errors smaller than 2 arcsec (with a 78 arcsec primary beam size) and therefore introduce negligible losses in the measured flux densities compared to other factors of uncertainty. Antenna temperatures were derived by applying a gain curve and opacity, which was measured at each observation session by tipping the antenna from horizon to zenith. Flux density accuracy is estimated to be $\sim$10 per cent.

Observations of the 5 x 5 arcmin region were conducted over two days; 2003 DOY 110 and 113. The region was surveyed in an 11 by 11 point grid with each point separated by 30" (a little less than the HPBW of 39") from adjacent positions. The correlator was configured to record 4096 spectral channels over a 16-MHz bandwidth for both left and right circularly polarised signals. At 22224 MHz this corresponds to a velocity coverage of $\sim$215 \kms and a spectral resolution of 0.06 \kmsns. Each point was observed for 30 seconds, resulting in an average RMS noise of 0.16~Jy, once both recorded circularly polarised signals were averaged together during data processing. 

Additional targeted observations towards seven bright regions (positions listed in Table~\ref{tab:bright24}) of 24-$\mu$m emission within the NGC346 region were carried out during 2008 DOY 200. These targets were selected on the basis of being potential young sites of high-mass star formation in the most active star formation region in the SMC \citep{Sabbi07}. During these observations a single circularly polarised signal was recorded across two 64-MHz bands centred on 22214 and 22244 MHz, resulting in continuous frequency coverage of 22182 to 22276 MHz. Across each band 2048 channels were recorded, resulting in a spectral resolution of 0.5~\kms over a velocity range of more than 1000~\kmsns. These observations were carried out as a series of two on-source observations bracketed by reference observations made 1$m$ earlier and 1$m$ later in right ascension. Each integration was one minute and the routine of on-source and reference observations was repeated four times resulting in a total on-source integration time of eight minutes per source, resulting in an RMS of $\sim$0.04~Jy.

\begin{table}
\begin{center}
  \caption{Locations of seven regions of bright 24-$\mu$m emission within NGC346 targeted for water maser emission with Tidbinbilla.}
  \begin{tabular}{cccccc}\hline\label{tab:bright24}
{\bf RA} &{\bf Dec} &&{\bf RA} &{\bf Dec}  \\
{\bf J2000}	&	{\bf J2000}	&&{\bf J2000}	&	{\bf J2000}\\
{\bf (h m s)}&{\bf ($^{o}$ $'$ $``$)} 	&&{\bf (h m s)}&{\bf ($^{o}$ $'$ $``$)} \\ \hline
00 59 09.1 &--72 11 00.3 && 00 59 01.4& --72 10 04.1\\
00 59 05.2 &--72 10 34.8 && 00 58 57.0 &--72 09 53.8\\
00 59 11.4 &--72 09 59.1 && 00 58 58.0 &--72 10 27.0\\
00 59 13.6 &--72 09 28.6 \\ \hline
	\end{tabular}
\end{center}
\end{table}

Monitoring observations of detected water masers were executed during short allocations of unused Host-Country time. Due to this, the observations varied from short $\sim$4 minute to much longer $\sim$16 minute on-source integrations, as reflected in the range of RMS noise values for each of the observations (listed in Table~\ref{tab:obs}). For the majority of these observations, a single circularly polarised signal was recorded across two 16 MHz bands with central frequencies separated by 5 MHz, resulting in a total bandwidth of 21 MHz (between 22215 and 22236 MHz). Across each 16 MHz band 4096 channels were recorded, corresponding to a velocity resolution of 0.06~\kmsns. 

Observations of two previously detected water masers in the SMC \citep{SB82}, s7 and s9, were made with Tidbinbilla on 2008 DOY 173. These observations used a slightly different setup, still recording a single circularly polarised signal, but with two 64-MHz bands. These bands were separated by 40 MHz and completely covered the frequency range from 22191 to 22295 MHz, corresponding to a velocity coverage of --550 to 700 \kmsns. In this configuration the spectral resolution is 0.5~\kmsns. 

Tidbinbilla data were reduced using the Australia Telescope National Facility (ATNF) Spectral Analysis Package ({\tt asap}). Alignment of velocity channels was carried out during the data reduction process, and all presented velocities are in the heliocentric reference frame. The adopted rest frequency for both these and the ATCA water maser observations was 22.23507985~GHz. The data have been corrected for the measured opacity of the atmosphere at the time of the observations. 

\subsection{ATCA water maser observations}

ATCA observations were made over 8 separate observing epochs, allowing us to determine accurate positions for the detected water masers (listed in Table~\ref{tab:detections}), as well as monitor their emission over time. All epochs, array configurations, spectral resolution, RMS noise and properties of the  water maser emission (if detected) are listed in Table~\ref{tab:obs}. 

The first epoch of ATCA data we present (2003 DOY 123) has been taken from the ATOA (project C973). These observations were made towards water masers detected in a search of 22 positions within the SMC, identified from {\em IRAS} and {\em MSX} colours to select potential \UCHII regions, oxygen-rich AGB stars or long-period variable stars. The observations were made in an EW352 array and the correlator was configured to record 512 spectral channels across a bandwidth of 16 MHz for two orthogonal linearly polarised signals. Target sources were observed in a series of 9, 3 minute ``normal'' scans, which in practice resulted in integration times between 2 minutes and 15 seconds to 2 minutes and 45 seconds per scan. Target observations were interspersed with observation of a phase calibrator, 0100--760. PKS~B1921--293 was observed as a bandpass calibrator, and primary calibration was made by fixing the flux density of PKS~B1921--293 to 12.015~Jy (to match measurements made in other observations at a similar time that were calibrated against PKS~B1934--638).

All six ATCA observations conducted in 2008 were made using either the H214 or 6B array configurations (details listed in Table~\ref{tab:obs}). During all 2008 observations, the correlator was configured to record 1024 channels across a 16 MHz bandwidth. In this configuration, only a single linearly polarised signal is recorded. All epochs included observations of either PKS~B1921--293 or PKS~B0537--044 for bandpass calibration and PKS~B1934--638 for primary flux density calibration. All observations were carried out as a series of cuts over the available hour angle range and were bracketed by observations of 0100--760 for phase calibration.

The 2011 ATCA observations were carried out over three consecutive days (total integration time of 12 minutes per source) and were able to take advantage of the capabilities offered by the new Compact Array Broad-band Backend \citep[CABB;][]{Wilson11}. With CABB we were able to record two orthogonal linear polarisations, with 2048 channels spread across a 64 MHz bandwidth. 

All epochs of ATCA data were reduced using the {\tt miriad} software package using the standard techniques for spectral line data. Image cubes of the entire primary beam and velocity range were typically created and inspected for each source. Spectra were created by integrating the emission in the imaged sources. In a minority of instances (specifically, observations of IRAS0043--7326 and IRAS01126--7332 on 2008 DOY 229) where we were unable to obtain sufficient $(u,v)$-coverage to make images, spectra were extracted from the vector averaged calibrated $(u,v)$-data. The RMS noise present in the spectra is typically better than $\sim$30~mJy depending on the individual integration times. The accuracy of the absolute flux density was checked by comparing the measured flux density of the phase calibrator which remained stable to within 10 per cent over the course of the observations. The positions of each of the detected sources reported in Table~\ref{tab:detections} are a weighted average position from at least five independent measurements for each source. The variations in the measured positions are typically small, and we expect that the reported positions are accurate to $\sim$0.5 arcsec.

\subsection{Targeted methanol and OH maser observations}\label{sect:other}

Observations of 1665-, 1667- and 6035-MHz OH and 6.7-GHz methanol maser emission were targeted towards the detected water masers during 2008. These observations were conducted using the Parkes 64 m radio telescope and the ATCA. The observation epochs, telescope, resultant 1-$\sigma$ sensitivity, spectral resolution and observed velocity range are listed in Table~\ref{tab:other_obs}.

ATCA observations of OH masers at 1665- and 1667-MHz, and 6.7-GHz methanol masers were made during an allocation of Director's time on 2008 DOY 203. These observations were interleaved with observations of the water masers also conducted at this epoch. Two frequency setups were used, one centred on 1664~MHz and the other at 6665~MHz. At both frequencies we recorded two orthogonal linear polarisations across a bandwidth of 4-MHz (allowing both the 1665- and 1667-MHz OH masers to be observed simultaneously) with 2048 spectral channels. For both frequencies, PKS~B1934--638 was observed  and used for primary flux density calibration, and observations of 2353--686 were used for phase calibration. Bandpass calibration was achieved from observations of PKS~B1934--638 at 1664 MHz and PKS~B1921--293 at 6665 MHz.

During the 2008 August session of the Parkes Methanol Multibeam Survey (MMB; see \citet{Green09} for a description) observations, we used some of the unallocated time after the Galactic Plane had set to target all four water masers for the presence of 6.7-GHz methanol, and 6035-MHz excited OH masers. These observations were carried out in the same manner as the MMB `MX' observations; a mode whereby a source is tracked and the pointing centre cycled through the receiver's beams. At the time of the observations, one of the 7-beamed receivers beams had a faulty LNA (low-noise amplifier) and was not used. Observations were carried out as a series of 10 minute integrations per beam, resulting in 60 minutes on-source for each observation. We repeated this process a further three times for NGC346~IR1 and twice for N66~Sc12, resulting in total integration times of four and three hours respectively.	The correlator recorded two frequency settings simultaneously, one centred on the methanol maser frequency, and the other on the 6035-MHz excited OH maser transition. For both settings, the correlator was configured to record two circularly polarised signals over a 4 MHz bandwidth with 2048 frequency channels.  

The OH and methanol ATCA data were reduced in the same manner as the ATCA water maser data; using the standard techniques for spectral line reduction in the {\tt miriad} software package. Images of the entire primary beam were created for each observed position, and RMS noise values were extracted at the positions of the water masers (and are reported in Table~\ref{tab:other_obs}). Parkes data were reduced using {\tt asap} as for the Tidbinbilla data.

\begin{table*}
\begin{center}
  \caption{Summary of targeted methanol and OH maser observations which failed to detect any maser emission. Column 1 lists the target water maser source, column 2 gives the observation epoch, column 3 lists the telescope (and array if the ATCA), column 4 lists the observed transition, and columns 5 - 7 give the 1-$\sigma$ sensitivity, spectral resolution and heliocentric velocity range of the observations.}
  \begin{tabular}{llllccccc}\hline\label{tab:other_obs}
 {\bf Source} &{\bf Date} & {\bf Telescope}& {\bf Maser}	&{\bf 1-$\sigma$} & {\bf Spectral} & {\bf Velocity} \\
		 & {\bf (DOY in 2008)}  & {\bf \& array} & {\bf transition}&{\bf sensitivity}& {\bf resolution} &{\bf coverage}  & \\ 
		 &			&			&	 &{\bf (mJy)}	&	{\bf (\kmsns)} &	{\bf (\kmsns)}\\ \hline

NGC346~IR1	&	203				&	ATCA, H214	&	1665-MHz OH	&	43		&	0.42	&	--100,600\\ 
				&			&		&	1667-MHz OH	&	43		&	0.42	&	250,950\\ 
				&				&		& 	6.7-GHz methanol 	&	144		&	0.11	&	70,250\\
				&236, 239, 242, 243		&	Parkes		&	6.7 GHz methanol	&	12		&	0.11	&	95,245				\\ 
				&	&		&	6035-MHz OH		&	17,15	&	0.12	&	90,260				\\ \\
				
N66 Sc12			&	203				&	ATCA, H214	&	1665-MHz OH	&	47		& 	0.42	&	--100,600		\\
				&					&				&	1667-MHz OH	&	47		& 	0.42	&	250,950	\\
				&					&				& 	6.7-GHz methanol 	&	150		&	0.11	&	70,250		\\
				&	234,	240, 241		&	Parkes		&	6.7 GHz methanol	&	12		&	0.11	&	95,245				\\ 
				&					&				&	6035-MHz OH		&	18,18	&	0.12	&	90,260		\\ \\

IRAS00430--7326	&	237				&	Parkes		&	6.7-GHz methanol	&	21		&	0.11	&	95,245	\\ 
				&					&				&	6035-MHz OH		&	36,32	&	0.12	&	90,260	\\ \\
				
IRAS01126--7332	&	238				&	Parkes		&	6.7-GHz methanol	&	23		&	0.11	&	95,245	\\ 
				&					&				&	6035-MHz OH		&	33,33	&	0.12	&	90,260	\\											
\hline
	\end{tabular}
\end{center}

\end{table*}

\section{Results}\label{sect:results}
\clearpage

Using the Tidbinbilla 70-m radio telescope and the ATCA we have discovered four water masers in the SMC. We have derived precise positions for each of these new water masers with the ATCA, and conducted several further observations over an eight year period to monitor the emission. Each of the observing epochs and details of detected emission at each observation are described in Table~\ref{tab:obs}. The characteristics of the detected water masers are listed in Table~\ref{tab:detections}, including: source name, equatorial coordinates, peak flux density and velocity and the minimum and maximum velocities detected at any of the observation epochs.

\begin{figure}
	\psfig{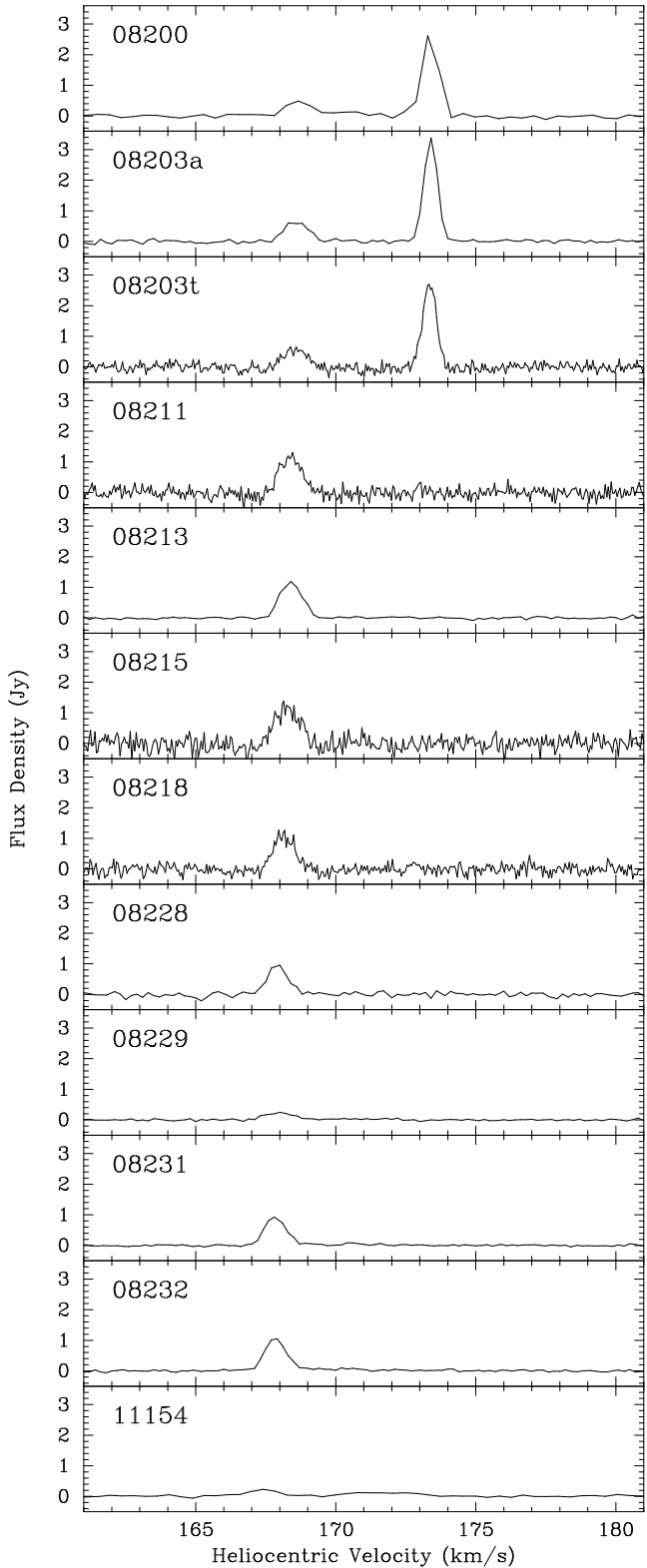}
\caption{Spectra of NGC346 IR1 at each of 12 observing epochs. In 2011 the peak flux density of the emission dropped to 0.24~Jy.}
\label{fig:smc_ir1}
\end{figure}

\begin{figure}
	\psfig{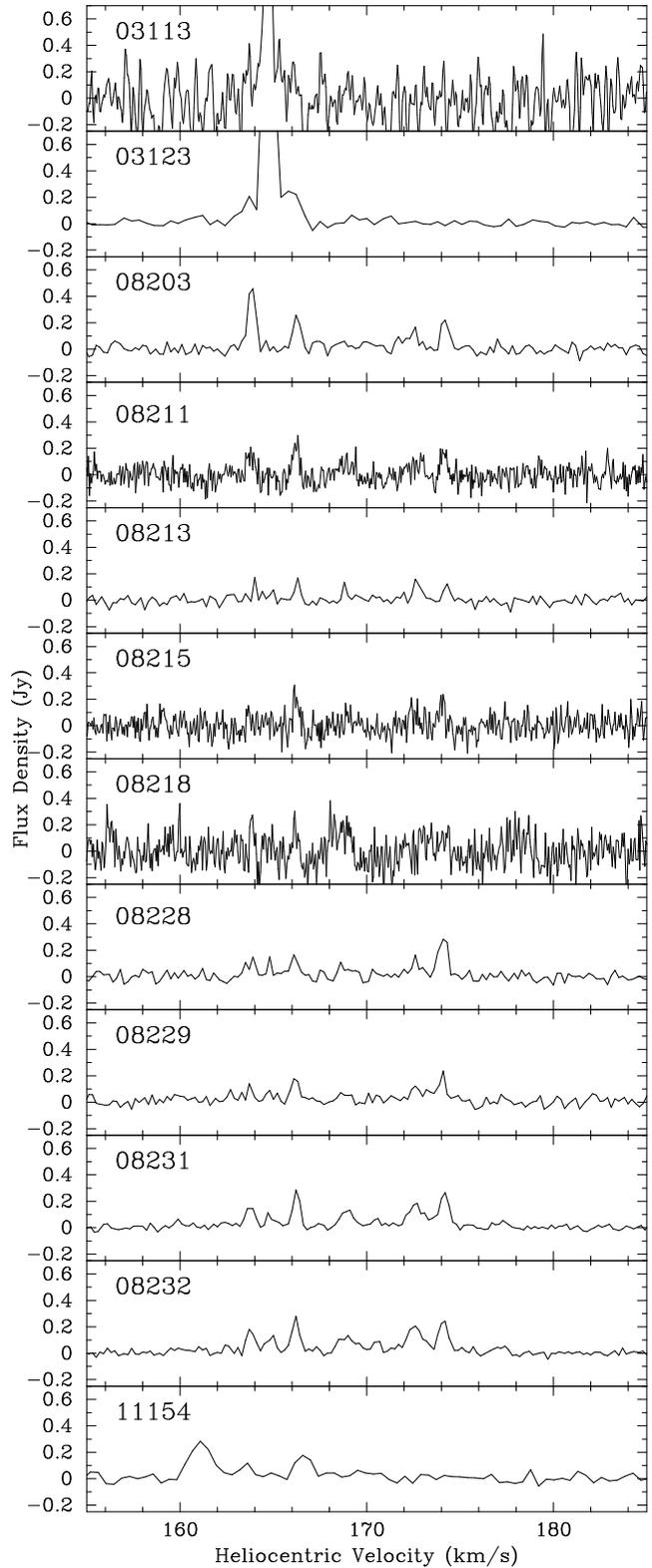}
\caption{Spectra of N66 Sc12 at each of 12 observing epochs. During both of the 2003 epochs the maser had a peak flux density of 1.6~Jy.}
\label{fig:lovell01}
\end{figure}

\begin{figure}
	\psfig{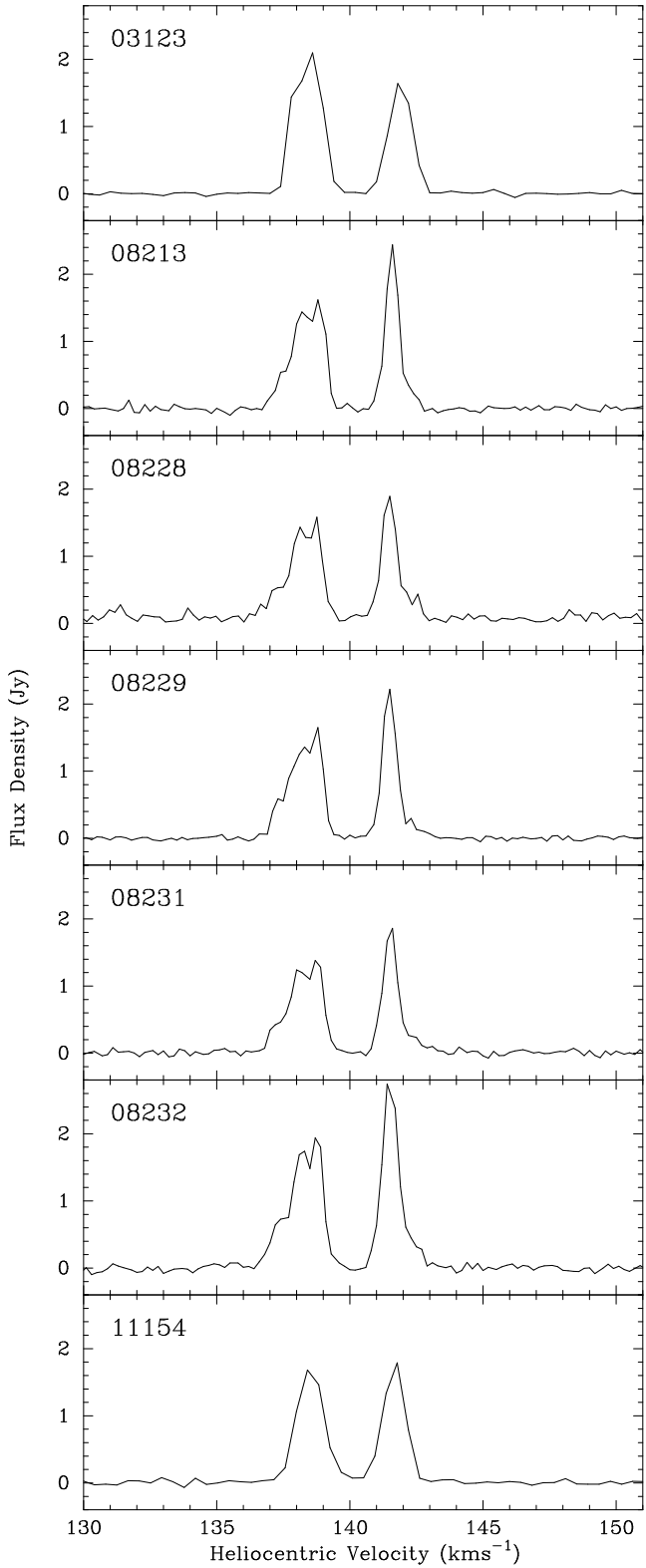}
\caption{Spectra of IRAS 00430--7326 at each of seven observing epochs.}
\label{fig:0043}
\end{figure}

\begin{figure}
	\psfig{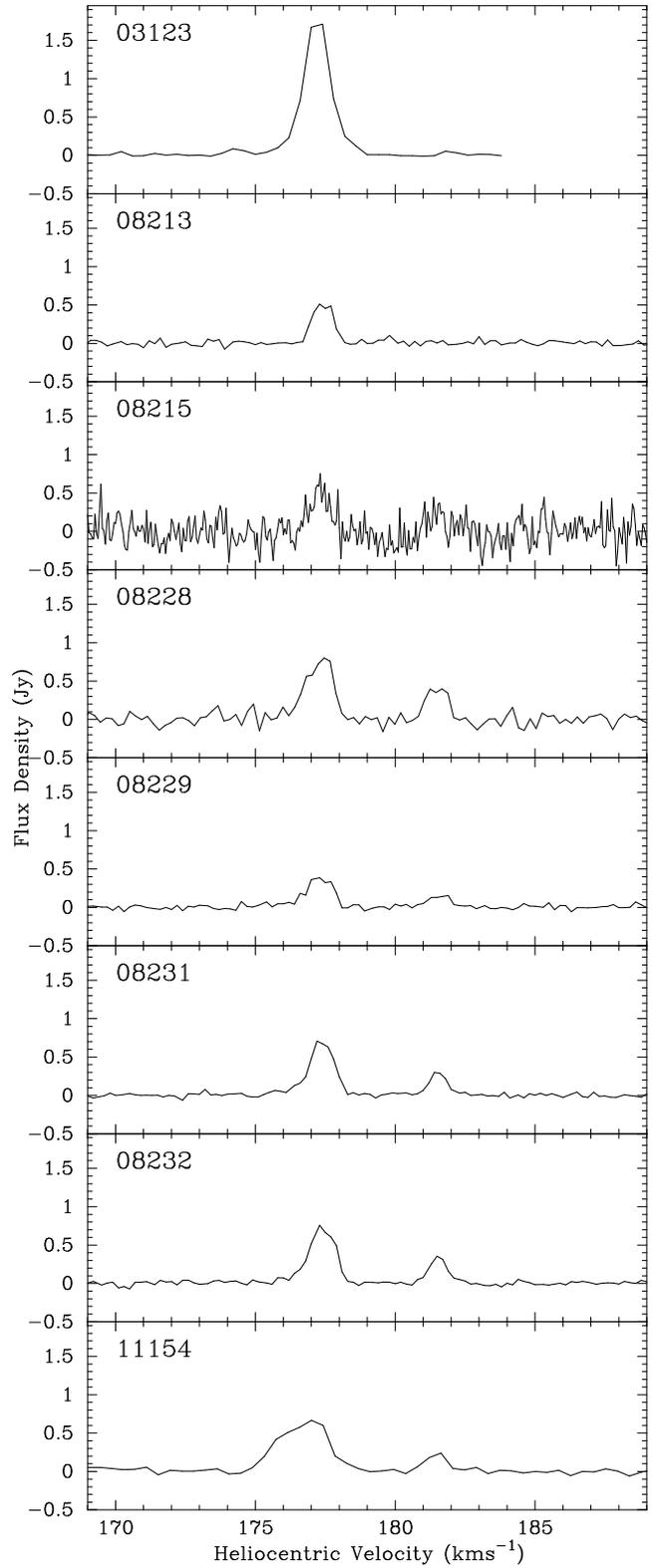}
\caption{Spectra of IRAS 01126--7332 at each of eight observing epochs.}
\label{fig:1126}
\end{figure}

Figures~\ref{fig:smc_ir1},~\ref{fig:lovell01},~\ref{fig:0043} and~\ref{fig:1126} present spectra for each of the detected water masers at each of the observing epochs. The date each spectrum was taken is annotated in the top left hand corner of each panel where the first two digits indicate the year and the following three, the day of year. Two spectra of NGC346~IR1 were taken on the same day and these are marked with either an `a' or a `t' following the observing date indicating observations with either the ATCA or Tidbinbilla, respectively.

Figure~\ref{fig:3col} shows the locations of the four water masers that we detect, as well as the two previously reported sources \citep{SB82}, overlaid on a three colour infrared image using data from the SAGE-SMC (Surveying the Agents of Galaxy Evolution in the Tidally Stripped, Low Metallicity Small Magellanic Cloud) {\em Spitzer} Legacy program \citep{Gordon11}. All four of the newly detected water masers are projected against bright regions of extended mid-infrared emission. This is perhaps not surprising given that our observations chiefly targeted regions on the basis of their infrared emission.

\begin{table*}
\begin{center}
  \caption{Summary of new water maser detections. Column one gives the source name; columns 2 and 3 give the average equatorial coordinates measured in our ATCA observations; columns 4 and 5 show the characteristics of the strongest spectral feature at any of the epochs (peak flux density and velocity); and column 6 gives the most extreme minimum and maximum velocities at any epoch. }
  \begin{tabular}{lllccccc}\hline\label{tab:detections}
{\bf Source}	&	\multicolumn{1}{c}{\bf RA} & \multicolumn{1}{c}{\bf Dec}  & 	{\bf Peak flux}	&	{\bf Vel peak}	& 	{\bf Vel range}\\
{\bf name}			&	\multicolumn{1}{c}{\bf J2000}	&	\multicolumn{1}{c}{\bf J2000}	& {\bf density}		&	{\bf (\kmsns)}	&	{\bf (\kmsns)}\\
				& \multicolumn{1}{c}{\bf (h m s)}&\multicolumn{1}{c}{\bf ($^{o}$ $'$ $``$)} 	&	{\bf (Jy)}\\ \hline
NGC346~IR1	&	00 59 09.31	&	--72 10 57.3	&	3.4	&	173.4	&	166.3,174.1 \\
N66 Sc12			&	00 59 19.70	&	--72 11 19.9	&	2.1	&	164.7	&	160.0,177.4	\\
IRAS00430--7326	&	00 44 56.44	&	--73 10 11.1	&	2.4	&	141.6	&	136.4,143.0		\\
IRAS01126--7332	&	01 14 04.74	&	--73 16 58.1	&	1.8	&	177.2	&	176.0,182.6\\ \hline
	\end{tabular}
\end{center}

\end{table*}

Follow-up observations for emission in the 1665-, 1667-MHz and 6035-MHz OH and 6.7-GHz methanol maser transitions were made with the ATCA and Parkes towards each of the four detected water masers. These observations failed to detect any emission from any of the observed transitions. The individual sensitivities of the observations are listed in Table~\ref{tab:other_obs}, and correspond to 5-$\sigma$ detection limits in the range 0.06 to 0.75~Jy.

\clearpage 

\clearpage

\section{Discussion}	

\subsection{Characteristics of the new maser sources}\label{sect:character}

The overall characteristics of our detected water masers are listed in Table~\ref{tab:detections}. The four sources have maximum peak flux densities in the relatively narrow range 1.8 - 3.4~Jy, and velocity ranges between 6.6 and 17.4~\kms (three of the sources lie between 6.6 and 7.8~\kmsns). Comparing these basic statistics with the sample of 379 Galactic water masers reported by \citet{Breen10} shows that the water masers in the SMC are not comparable to highest luminosity Galactic sources, and show much narrower velocity ranges. Of the 379 Galactic sources, 36 have flux densities between 180 and 1600~Jy, which if we assume are located at a distance of 6 kpc, would be detected with flux densities between 1.8 and 16~Jy if placed at the distance of the SMC. The average velocity range of emission features greater than 20~Jy (meaning they would be detectable in most of our observations if placed at the distance of the SMC) of these 36 sources is $\sim$25~\kmsns, with some sources showing features that would be detected across more then 100~\kmsns. Neither the current SMC sample or the Galactic sample of \citet{Breen10} are representative of the complete water maser populations given their inherent biases (more than half of the \citet{Breen10} sample are associated with their OH or methanol masers targets) and more thorough comparisons can be made once recent complete Galactic water maser surveys \citep[e.g.][]{Walsh11} have been reported at high angular resolutions. However, it is likely that surveys such as that of \citet{Breen10} much better represent the water masers with the higher flux densities (as considered here) than the weaker water maser sources which are more often solitary \citep[e.g.][]{Breen10,BE11}.

Water masers associated with star formation regions in our own Galaxy are often accompanied by masers of other species, such as OH and methanol \citep{FC89,Beuther2002,Breen10}. The SMC has been completely searched for both 6.7- GHz methanol and 6035-MHz OH masers \citet[as part of the MMB survey: ][]{Green08} resulting in no detections to a 1-$\sigma$ detection limit of 0.13~Jy. Although we made more sensitive observations at the frequencies of 6.7-GHz methanol masers and several OH maser transitions, we also failed to detect any emission. \citet{Breen10} note that their stronger water masers are more often associated with either or both OH and methanol masers, seemingly in contrast to the SMC sources. To determine if this difference is significant, we have considered the large maser samples of \citet{Breen10} and \citet{C98,Cas03,C09}. 

Table~\ref{tab:detections} shows that all of the SMC water masers have been at least 1.8~Jy on at least one of the observing epochs, the equivalent of a 180~Jy Galactic source. \citet{Breen10} detect 36 water masers in our Galaxy with a flux density of $>$180~Jy and of these only three are not associated with either OH or methanol masers. Of these 36 water masers, 17 are associated with methanol masers with peak flux densities $>$10~Jy (14 of the 17 are $>$20~Jy) which would be equivalent to a $>$0.1~Jy source at the distance of the SMC and therefore detectable in our observations. Similarly, 10 of the 36 water masers are associated with OH masers with peak flux densities greater than 25~Jy that would correspond to 0.25~Jy in the SMC, which is $\sim$5 times our 1-$\sigma$ detection limit as listed in Table~\ref{tab:other_obs}. Only one of the strong Galactic water masers is associated with a 6035-MHz excited OH maser that it would be strong enough to be detected if at the distance of the SMC.

Given that 47 per cent of equivalent Galactic water masers have associated methanol masers strong enough to be detected by our observations at the distance of the SMC, it may be that the characteristics of the four SMC sources are inconsistent with those in our Galaxy. However, it is not surprising that we fail to detect any OH emission from the two sources targeted for 1665- and 1667-MHz transitions given only 28 per cent of similar Galactic water masers are associated with 1665- or 1667-MHz OH masers that would be detected if placed at 60 kpc. Similarly, we would not expect to detect 6035-MHz OH masers towards the SMC water masers if their properties are similar to Galactic sources.

\citet{Green08} found that methanol masers in the LMC were under abundant by a factor of 4 - 5 once the different star formation rates had been accounted for. This under abundance was attributed to lower metallicity in the LMC compared to out Galaxy, in agreement with the earlier study of \citet{Beasley96}. The SMC similarly has a lower metallicity than our Galaxy \citep[$\sim$one-fifth solar;][]{Peimbert00} so it is expected to also show a significant under abundance of methanol masers which may explain why neither our sensitive targeted observations, or the complete search of \citet{Green08}, uncovered any emission from this transition.

\subsection{Observations of previously identified water masers in the SMC}

Observations targeting the two water maser positions reported by \citet{SB82} were made on three separate occasions; twice with the ATCA (03123 and 08213) and once with Tidbinbilla (08173), and failed to detect any emission towards either source to 5-$\sigma$ detection limits in the range 75 to 175~mJy over the minimum observed velocity range of 10 to 220 \kmsns. When detected more than 30 years ago, s7 peaked at 7.4~Jy at a velocity of 120.8 \kms and s9 exhibited two spectral features, the strongest of which was 4.2~Jy at a velocity of 120.9 \kmsns. Our observations show no emission at these, or nearby velocities. 

Since the HPBW of the earlier observations was 4.5 arcmin, it is possible that the sources fell outside our much smaller beams. However, given the sensitivity of our observations, it is likely that we would have detected them anyway if they remained at flux densities similar to those reported by \citet{SB82}, especially given the poor spectral resolution of their observations (1.35~\kmsns) which could have significantly underestimated the flux density of these sources. 

\begin{figure*}
  \begin{center}
	\psfig{figure=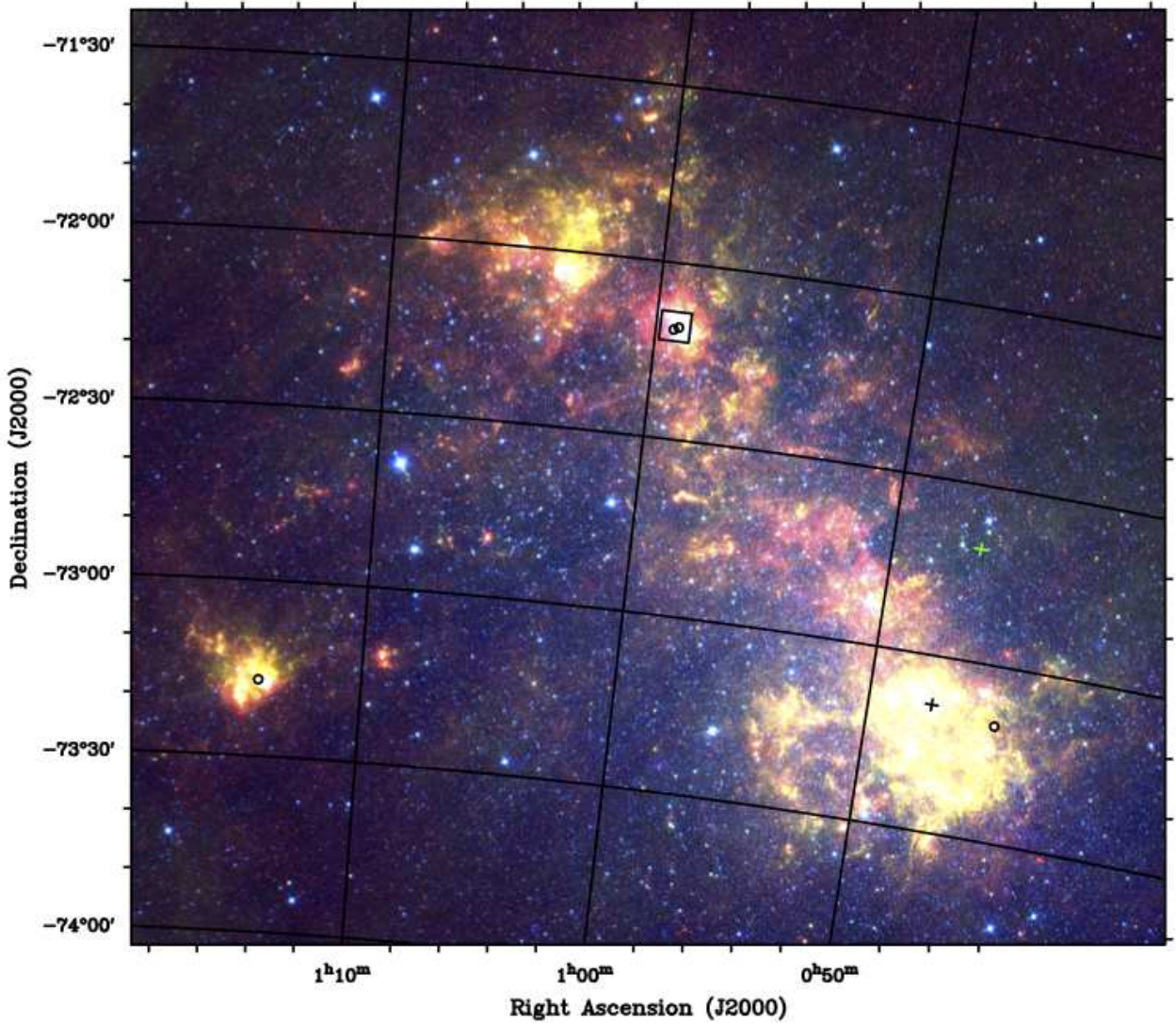,width=17cm}
\caption{Three colour image of the SMC (red=24-$\mu$m MIPS emission, green=8.0-$\mu$m IRAC emission and blue=4.5-$\mu$m IRAC emission from the SAGE-SMC {\em Spitzer} Legacy program \citep{Gordon11}). The positions of the four water masers we detected (circles) and previously detected water masers (marked by green (s7) and black (s9) crosses; \citet{SB82})  are overlaid. The border of our 5 $\times$ 5 arcminute survey region is also marked.}
\label{fig:3col}
\end{center}
\end{figure*}

Fig.~\ref{fig:3col} shows the locations of s7 and s9 overlaid on a three colour infrared image. It can be seen that s9 (black cross) is projected against a bright region of infrared emission, while s7 (green cross) is on the periphery of the bright emission. Looking at the spectra presented in \citet{SB82}, it is s7 that has the higher signal-to-noise, lending credence to its authenticity even though it is located in a seemingly unlikely location with respect to the infrared emission. In order to rule out any persistent emission, sensitive mosaic observations centred on the reported positions need to be conducted. 

Given that the observations were conducted more than 30 years ago, it is also possible that the emission has genuinely dropped below detectable limits. \citet{Felli07} conducted monitoring observations of 43 water masers, associated with young stellar objects (YSOs), over a 20 year period. Their observations show that it is not unusual for water masers to vary on relatively short timescales, and in fact show that a number of their sources fall below their detection limit on several occasions, generally returning or replaced by emission at a similar velocity sometime later. However, this characteristic is not sufficiently common to explain the appearance and disappearance of the only two previously detected water masers in an approximately synchronised  fashion. The more likely scenario seems to be that the reported positions are sufficiently poor that we missed one or both of them with our much smaller beams, or perhaps less likely given the convincing spectra and their locations with respect to infrared emission, one or both were never authentic detections.

\clearpage

\subsection{Variability of detected water maser emission}\label{sect:var}

Water maser emission in our Galaxy is known for exhibiting much higher levels of temporal variability over relatively short time-scales than other commonly observed maser species \citep[e.g.][]{Felli07}. Our water maser observations span a number of years, with varying sampling rates, allowing us to assess the overall level of variability in the water masers we detect, and determine whether this is consistent with Galactic sources. In the following paragraphs we discuss the variability of each of the detected masers and then compare their variability with similar Galactic sources.

{\em NGC346~IR1} was observed once in 2003, 11 times in 2008 and once in 2011. No emission was detected during the 2003 observations to a 5-$\sigma$ detection limit of 0.29~Jy. Then, in 2008 we detected two relatively strong features, the strongest of which ($\sim$3~Jy) was observed over just three days, disappearing during the eight days between our 08203 and 08211 epochs. This feature was not detected in any further observations, so its lifetime appears to be less than five years. The spectral feature at $\sim$168 \kms has basically remained stable (aside from the changes in velocity discussed in Section~\ref{sect:acc}) over the course of the observations in 2008 (11 epochs over 32 days). Observations in 2011 revealed a weaker feature at a similar velocity which is inconsistent with the velocity that would be expected if the feature continued to accelerate at the rate seen in 2008. It is therefore possible that this component has been replaced by similar emission. 

On day 08203 we observed this source with both the ATCA and Tidbinbilla. The measured peak flux densities of the stronger feature at $\sim$173~\kms were 3.4 and 2.7~Jy at the ATCA and Tidbinbilla, respectively. These values are marginally inconsistent with our 10 per cent calibrations errors. It can be seen that the secondary feature in both observations is approximately stable at $\sim$0.6~Jy. While it is possible that there is some real variability seen in this source (although the time range of the observations are similar, the ATCA spectrum is the average value over several hours, whereas the Tidbnbilla spectrum is a small snapshot of this time), it is likely that the measured flux density difference can largely be accounted for by the different noise contributions and spectral resolutions in the respective observations.

{\em N66~Sc12} has the most complicated spectral structure of the detected sources, exhibiting many features over the broadest velocity range (17.4 \kmsns). It is also the weakest source during many of the epochs making a quantitative study of its variability difficult. In 2003 it showed essentially a single, strong (1.6~Jy) spectral feature that was not detected in any of our 2008 or 2011 observations. Instead, during the later epochs the emission was replaced by a plethora of weaker features, showing few components that can be tracked over significant time ranges. Although the signal-to-noise of many of the observations conducted in 2008 is low, it appears that some of the features associated with this source show variability on time scales of days.

{\em IRAS00430--7326} shows the most consistent and stable emission of all the SMC water masers. Given the relatively poorer spectral resolution during the single 2003 and 2011 observations, it is difficult to comment on the stability of the individual components of the clearly complicated and blended structure over the full time period, but the strongest features are clearly persistent. 

{\em IRAS01126--7332} shows moderate levels of variability. The component at $\sim$177.5~\kms first detected in 2003 persists all the way through to our final observation in 2011 (although reduces in flux density by more than half), and exhibits an accompanying feature from 08228 at $\sim$181.5 \kmsns. The final spectrum shows a broadening in the original feature at a velocity of $\sim$177.5~\kmsns, perhaps indicating the presence of a new feature that is spectrally blended with the persistent one. 

A number of the detected sources show levels of variability that would be rated amongst the most variable star formation masers in our Galaxy \citep[e.g.][]{Felli07}. For example, the disappearance of the 3~Jy feature of NGC346~IR1 on a time scale of a week is equivalent to a $\sim$300~Jy source reducing to less than $\sim$12~Jy. \citet{Breen10} highlighted one of their 379 Galactic star formation water maser sources as showing extreme variability; fading from 80~Jy  to $<$0.2~Jy in a period of less than a year, a reduction that is probably less extreme than the feature associated with NGC346~IR1 (given the relative timescales). The appearance of the feature in the 08228 observations of IRAS01126--7332 would be equivalent to a $<$3~Jy source increasing by a factor of $\sim$10 over a period less than two weeks, similarly showing quite extreme variability. Arguably the most variable source in the SMC is N66~Sc12, showing evidence of significant change over a matter of days, rarely showing the same spectral structure from epoch to epoch. In addition, the probable variation of the two out of two previously detected water masers \citep{SB82} over 30 years would be an unlikely event in equivalent Galactic sources of more than $\sim$400~Jy.  

Due to the small sample size, it is difficult to draw definitive conclusions about the global variability properties within the SMC, however, the characteristics of the detected sources show evidence of higher levels of variability (on average) on very short timescales than similar Galactic sources, perhaps consistent with them being influenced by a generally more dynamic environment. Future monitoring observations of the water masers in the LMC may also provide a useful comparison.

 \subsection{Accelerating feature associated with NGC346~IR1}\label{sect:acc}
 
\begin{figure}
	\psfig{figure=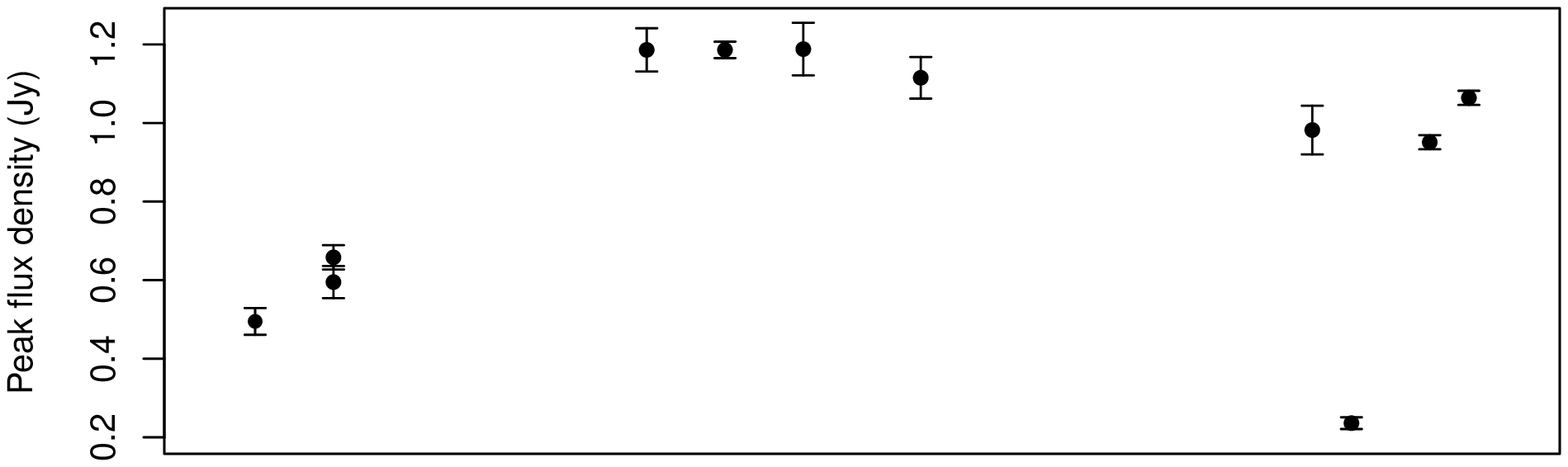,width=4.5cm, angle=270}\vspace{-2.1cm}
	\psfig{figure=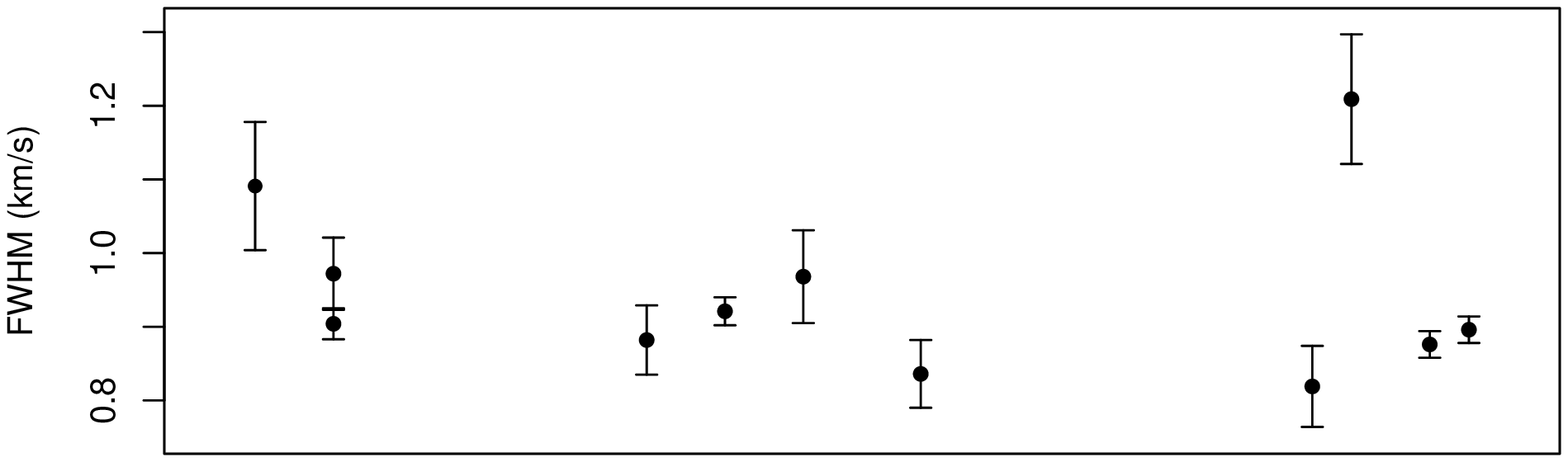,width=4.5cm, angle=270}\vspace{-2.1cm}
	\psfig{figure=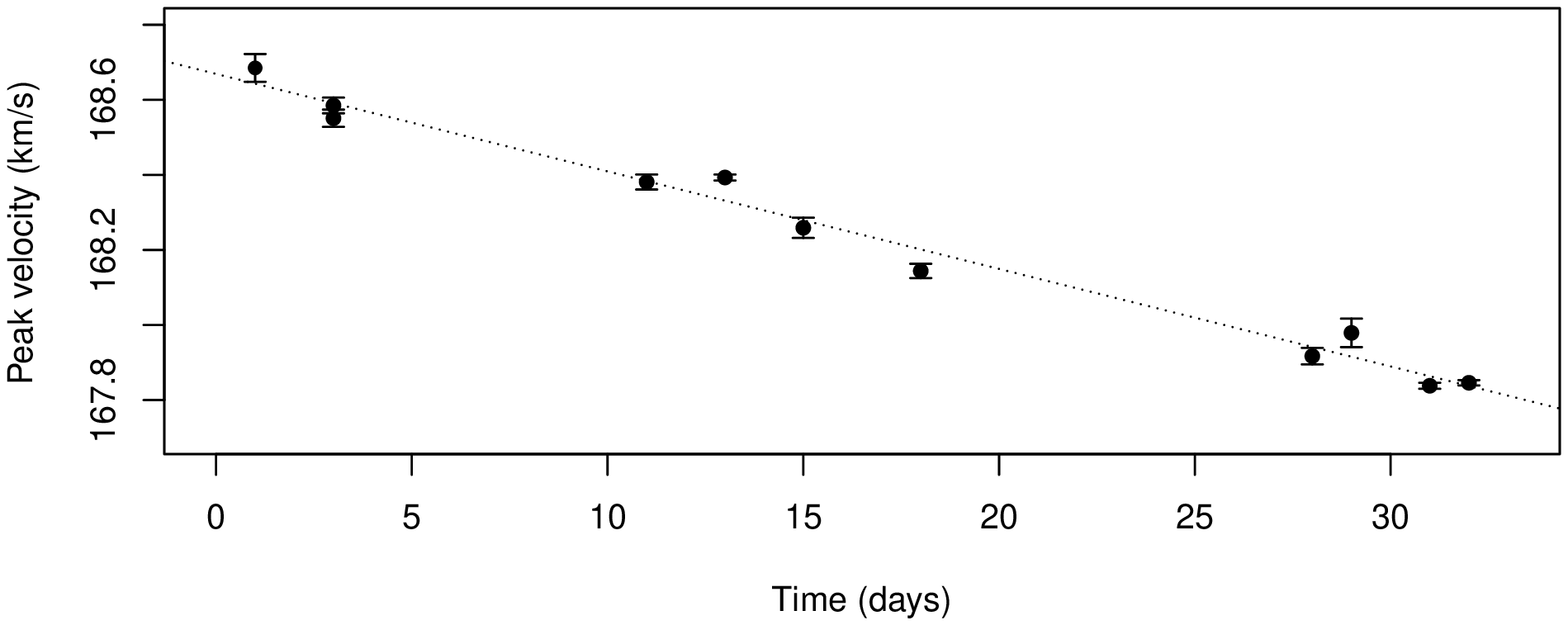,width=4.5cm, angle=270}
\caption{Peak flux density (Jy), FWHM (\kmsns) and peak velocity (\kmsns) of the persistent feature at $\sim$168~\kms associated with NGC346~IR1. Overlaid on the peak velocity plot is the line of best fit to the data. All parameters are derived from Gaussian fits to the emission and the error bars represent the formal errors in the Gaussian fit. All data collected in 2008 are plotted, totalling 12 epochs (including two on the same day). Here time 0 days corresponds to 2008 DOY 200.}
\label{fig:ngc}
\end{figure}

 The spectra of NGC346~IR1 presented in Fig.~\ref{fig:smc_ir1} show an apparent velocity shift in the feature at $\sim$168~\kmsns, tending to less positive velocities as time progresses. Fig.~\ref{fig:ngc} shows the peak flux density, line FWHM and peak velocity parameters from Gaussian fits to the detected line during each of the observations carried out in 2008. Also overlaid on the peak velocity plot is the line of best fit, with parameters of intercept = 168.68 and slope = --0.027; corresponding to a drift in velocity of 9.6 \kms yr$^{-1}$ over the 31 day period.  As can be seen in Fig.~\ref{fig:ngc} there is no correlation between the change in peak velocity with either peak flux density or the FWHM of the detected emission, ruling out the possibility that the change in velocity is influenced by the variability of a spectrally blended feature.
 
 A number of studies have noticed similar behaviour in water maser components associated with Galactic high-mass star formation regions. The fraction of water masers that show velocity drifts is probably grossly underestimated due to the compounding effects of complex spectra, variability and the small number of sources that have been monitored. In spite of this, \citet{Brand03} reported 15 isolated maser components with accelerations in the range 0.02 to 1.8~\kms yr$^{-1}$ (with equal numbers of positive and negative acceleration) over time spans of 60 to 4600 days. Similarly, \citet{Hunter94} noted an outward acceleration in one of the water maser features associated with W75N of 1.45 \kms yr$^{-1}$, while \citet{Motogi11} found that two of the spectral feature associated with G\,353.273+0.641 underwent synchronised changes in velocity at a rate of $\sim$--5 \kms yr$^{-1}$, occurring during a maser flaring period. \citet{Cas04} also noted acceleration in the blue-shifted emission of two sources associated with Galactic star formation: G\,291.270--0.719 and G\,291.284--0.716. The changes in velocity were noted over long periods (more than 20 years) and correspond to accelerations of 0.7 and 0.4 \kms yr$^{-1}$, respectively. \citet{Cas04} concluded that the accelerating features were likely associated with outflow material that is still fairly close to its origin (otherwise it would be expected to begin decelerating). \citet{Brand03} offer additional possible explanations for observed velocity drifts, including the rotation of a disk, changes in the velocity of a collimated outflow or the precession of an outflow.

Given that the systemic velocity for this source is not known to high accuracy \citep[nearby CO clouds show comparable velocities;][]{Muller10}, we are unable to determine its positive or negative acceleration, or how close this feature is to the systematic velocity, making physical interpretation difficult. Our acceleration is larger than any of the reported values for Galactic water masers \citep[although similar to some extragalactic water masers; e.g.][]{Greenhill95} and is also the smallest time range considered. In general, there seems to be a trend in the reported values of acceleration to be smaller when measured over longer time periods. Our relatively high acceleration perhaps indicates that the emission is associated with outflow material that is very close to its origin. An alternative scenario might be that the sparseness of the observations and longer time period considered in some sources has masked bursts of acceleration that have subsequently been averaged out in the calculated accelerations, which may also be consistent with a precessing outflow. 

We conducted one further observation of NGC346~IR1 in 2011 (not included in Fig.~\ref{fig:ngc}) that is not consistent with that expected from the simple linear fit to the 2008 data. If we include these data and determine the average acceleration over the $\sim$3 year period, it drops to $\sim$0.5 \kms yr$^{-1}$, perhaps indicating that over a longer period the characteristics of this source are not unlike other Galactic sources presented in the literature. 
However, our favoured explanation is that the feature observed in 2008 disappeared in the intervening period before our 2011 observations (perhaps due to disruptions caused by an outflow) and was replaced by generally similar emission at a slightly different velocity. This scenario is supported by the fact that the accompanying characteristics also exhibited more extreme values (peak flux density = 0.22~Jy; FWHM = 1.27 \kmsns) than the range seen in the 2008 observations.

%

\subsection{Using the water masers for proper motion studies}

The Magellanic Clouds are the focus of detailed studies for a number of reasons, one of which is that they are clearly interacting, both with each other and with the halo of the Milky Way.  Evidence for this is seen particularly in the H{\sc i} distribution which shows tidal features - the Magellanic Stream, the Leading Arm and the Magellanic Bridge \citep[see e.g.][]{McClure-Griffiths+09}.  Interactions and mergers of galaxies are known to play a key role in their evolution throughout cosmic history, and the Magellanic system provides an opportunity for the key factors and effects of interactions to be investigated at high spatial resolution and sensitivity.  Our understanding of the way in which the Magellanic Clouds are interacting with the Milky Way has changed dramatically in the last decade due to new measurements of their proper motions \citep{Kallivayalil+06,Piatek+08}, which suggest that they are not gravitationally bound to the Milky Way (as has previously been thought), and are undergoing their first (and last) interaction with it \citep{Besla+07}.  The measurements of the proper motions of the LMC and SMC based on {\em HST} studies of a number of fields in both galaxies remain controversial.  Dynamical simulations which attempt to model the distribution of stars and gas in the Magellanic system using these proper motions have been unable to reproduce observed structures such as the bifurcation of the Stream and Leading Arm \citep{Diaz12}.

One of the complications in the existing optical proper motion studies is that they estimate the proper motion by averaging over all the stars in a particular field.  However, the SMC has a significant line-of-sight depth which means that any field contains contributions from more than one dynamical population - both the halo and disk of the galaxy.  The uncertainty due to this is not included in the formal error estimates of the LMC and SMC proper motions determined from the {\em HST} data \citep[see][]{Diaz12}.  The water masers we have discovered in the SMC are both strong (peak flux density greater than $\sim$1~Jy), and compact (angular scales less than 1~milliarcsecond), which means that they are ideal targets for VLBI. 

VLBI observations have achieved astrometric accuracies better than 10~$\mu$as for a single epoch with instruments such as the VLBA (Very Long Baseline Array) and VERA (VLBI Exploration of Radio Astronomy) \citep[e.g.][]{Reid09}.  Indeed, the most accurate measurements of proper motions of northern local group galaxies M33 and IC10 have been from measurements of water masers in these systems \citep{Greenhill93,Brunthaler+05,Brunthaler+07}.  The proper motion of the SMC estimated from {\em HST} data is $\mu_{\alpha} = 0.75 \pm 0.06$ mas/year ; $\mu_{\delta} = -1.25 \pm 0.06$ mas/year \citep{Piatek+08}.  So it should be possible to measure the proper motion of the SMC to an accuracy better than 10 percent using VLBI observations with a temporal baseline of 1--2 years.  The primary difficulty in making proper motion observations of water masers is that they exhibit significant temporal variability (see Section~\ref{sect:var}), making it difficult to track individual maser component ``spots'' over the course of an extended VLBI program.

The water masers in the SMC are all associated with young, high-mass star formation regions, which means that they are found in regions of dense gas.  This gas shows much lower velocity dispersion than the general stellar populations, and as the masers are spectral line sources, the combination of a proper motion observation with the line-of-sight velocity gives the full three dimensional velocity vector of the object.  However, although water masers offer some benefits for proper motion studies compared to those based on stellar populations, they have their own shortcomings and sources of uncertainty.  The observed proper motions of interstellar masers in the SMC will contain three contributions : 1. A component due to the proper motion of the centre of mass (CoM) of the SMC ; 2. A component due to the orbital motion of the star formation region about the CoM of the SMC ;  3. A component due to internal motions of the masing gas within the star formation region.  The first of these is the critical measurement, but the uncertainty in any determination of this depends upon both the precision in the proper motion measurement and the accuracy with which the other two components can be estimated.  Water masers in star formation regions typically arise in shocks or outflows, and velocities in excess of 100~\kms\/ are not uncommon in Galactic water masers \citep{Breen10,BE11}.  At an assumed distance of 61~kpc, a velocity uncertainty of $\sim$10~\kms\/ corresponds to a proper motion uncertainty of $\sim$0.04~milliarcseconds per year (or a few percent).  In addition to this, with only four water masers known in the SMC and variability potentially making it challenging to obtain proper motion measurements from some of these, uncertainties due to random errors will dominate (unlike the {\em HST} measurements where it is systematic effects which are more important).

In summary, astrometric VLBI observations of the water masers in the SMC offer a potential alternative means of measuring its proper motion. Such measurements could be achieved, if made with careful calibration of the residual path lengths, with the current capabilities of the Australian Long Baseline Array for water maser sources with peak flux densities less than 2.5~Jy when there is a relatively strong ($>$100 mJy) calibrator within 1-2 degrees of the targets. These measurements are free from some of the systematic errors which potentially affect the optically-based proper motion determinations.  Although they have their own sources of uncertainty, they do provide an independent means of determining the proper motion of the SMC.

\subsection{SED modelling}

\citet{Ellingsen2010} investigated the properties of the infrared YSOs associated with the LMC water masers and compared the properties of their spectral energy distributions (SEDs) with those of the general population of YSOs in the LMC identified by \citet{Gruendl+09}.  They showed that the maser-associated YSOs were more luminous and have redder infrared colours than the population as a whole and that the SEDs inferred the maser sources have higher mass, luminosity and ambient density than the majority of YSOs in the LMC.  

We have used the results from the {\em Spitzer} SAGE-SMC project \citep{Gordon11} to investigate the YSOs associated with SMC water masers for which we have measured the position with the ATCA.  The SAGE-SMC observations have angular resolution of around a few arcseconds for the observations at wavelengths less than 10~$\mu$m and a few tens of arcseconds for wavelengths greater than 20~$\mu$m.  The positions of the infrared sources are accurate to better than 1 arcsecond, so investigation of the properties of the S7 and S9 maser candidates cannot be undertaken, since the position for these sources is uncertain at the arcminute level.  All four of the SMC water masers detected in the current observations have a SAGE-SMC source with 0.5 arcseconds of the maser position. \citet{Oliveira+13} identified 34 candidate YSO in the SMC which are bright at both 24- and 70-$\mu$m and undertook mid-infrared spectroscopy of these sources.  IRAS00430--7326 is the strongest source at 70~$\mu$m in the \citet{Oliveira+13} sample (source \#03 in their list) and from their spectra they identify polycyclic-aromatic hydrocarbon (PAH) emission, silicate absorption and both water and CO$_2$ ice absorption.  The presence of ice absorption indicates that the sources are surrounded by a cool envelope, and hence highly embedded and young sources.  Of the other three SMC water masers IRAS01126--7332 has a bright 70-$\mu$m source offset from the maser position by a little over 2 arcseconds, but the remaining sources do not have a nearby 70-$\mu$m point source (and hence were not included in the \citeauthor{Oliveira+13} sample).

\begin{figure*}
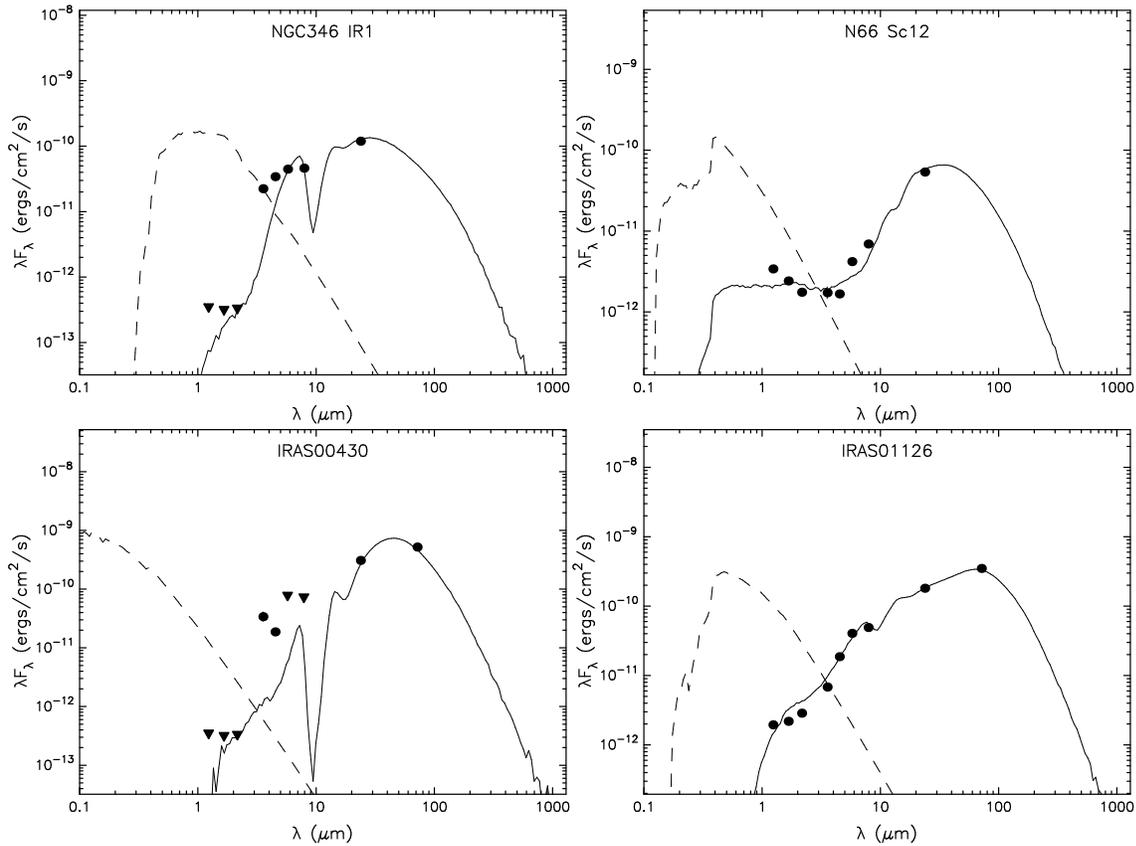

	\psfig{figure=ngc346.eps,width=7.5cm}\psfig{figure=N66.eps,width=7.5cm}
	\psfig{figure=iras_00430_carlson.eps,width=7.5cm}\psfig{figure=iras_01126.eps,width=7.5cm}
\caption{The SED and fitted models for the sources associated with the four detected water maser sources in the SMC. The dots and the triangles are measurements and upper limits taken from the 2MASS and SAGE-SMC surveys. The solid line shows the best fitting model, while the dashed line gives the spectrum of the stellar photosphere in the absence of circumstellar dust but accounting for interstellar extinction. }
\label{fig:sed}
\end{figure*}

Using the online SED fitting tool of \citet{Robitaille+07}, we have investigated the SEDs of the maser-associated YSOs in the SMC.  We have used the same general approach as applied to LMC YSOs by \citet{Ellingsen2010}, the only differences being that for the SMC we allowed the fitted distance to vary between 59 and 62~kpc, and for sources where there is strong evidence for PAH emission contaminating the mid-infrared photometry we used the method of \citet{Carlson+12}, to mitigate its effects on the SED fitting.  The YSO models of \citet{Robitaille+06} (which are used in the SED fitting), do not include the affects of PAH emission, which can significantly affect the measured intensity in the 3.6-, 5.8- and 8.0-$\mu$m {\em Spitzer} bands.  The only {\em Spitzer} IRAC band which is not affected by PAH emission (when it is present) is the 4.5-$\mu$m band, and \citet{Carlson+11} identified sources with a characteristic dip in this band using the criteria [3.6]-[4.5] $<$ 0.5 and [4.5]-[5.8] $>$ 1.5, as those with significant PAH emission.  Figure~9 of \citet{Ellingsen2010} shows that this characteristic dip at 4.5-$\mu$m is commonly seen in maser-associated YSO in the LMC and we find that all of the maser-associated YSO in the SMC show a dip at 4.5-$\mu$m in the SED plots (see Figure~\ref{fig:sed}).  The only source for which the PAH emission is significant (using the criteria of \citeauthor{Carlson+11}) is IRAS00430--7326.  For this source we used the approach outlined in \citet{Carlson+12} for sources with a 24-$\mu$m measurement (this involves setting the measured 5.8- and 8.0-$\mu$m intensities as upper limits and increasing the estimated uncertainty at 3.6-$\mu$m to 20 percent).  For all other sources we used the 2MASS and SAGE-SMC intensities and estimated uncertainties in our SED fitting.

The results of the SED fitting can be seen in Figure~\ref{fig:sed}.  They show that the infrared observations for all sources are consistent with the masers being associated with deeply embedded, highly luminous YSOs. \citet{Ellingsen2010} found significant differences between the maser-associated YSO and the rest of the YSO population in the mass of the central source, the total luminosity and the ambient density (see figure 10 of \citeauthor{Ellingsen2010}).  Comparing the results obtained for the SMC maser-associated YSO with those in the LMC we find that they fall in the same range, with the mass of the central source greater than 15~\Msol, the total luminosity greater than 15 000~\Lsol\ and the ambient density greater than $4.5 \times 10^{-21}$ cm$^3$ in all four sources.  This strong similarity is to be expected, and suggests that future water maser searches targeting SAGE-SMC sources with characteristics similar to those of the known maser-associated sources are likely to find additional masers in the SMC.

\section{Conclusion}

From a number of targeted water maser observations conducted with both the Tidbinbilla 70-m radio telescope and the ATCA we present the discovery of four water masers in the SMC. In a series of sensitive observations, we fail to detect any water maser emission towards the positions reported in \citet{SB82}. Our monitoring observations of each of the detected water masers, combined with the non-detection of the previously detected sources, show evidence for the SMC water masers to be more variable, in general, than similar Galactic sources also associated with YSOs. These observations also revealed that one of our water masers (NGC346~IR1) exhibited an accelerating feature at a rate of 9.6~\kms yr$^{-1}$.

We conducted targeted observations for methanol and OH masers at the positions of the newly detected water masers with both the ATCA and the Parkes 64-m radio telescope and found no emission from any of the observed transitions. The absence of methanol maser detections in our targeted observations, combined with the results of the complete survey of \citet{Green08}, is consistent with the SMC showing an under abundance of methanol masers compared to our Galaxy, likely due to lower metallicity \citep[as was found for the LMC;][]{Green08,Beasley96}.

The SEDs of the YSOs associated with our detected water masers have been investigated and reveal that the sources associated with water masers in the SMC are comparable to those in the LMC \citep{Ellingsen2010}. It is expected that targeted water maser observations towards sources with similar characteristics will be a successful criterion for finding more water masers in both Magellanic Clouds.  

The detected water masers will make excellent targets for future VLBI observations to provide and independent, accurate measurement of the proper motions of the SMC. Our monitoring observations have shown that such a program will need to have a relatively high cadence to accurately track individual components used to measure the proper motion of the galaxy.

\section*{Acknowledgments}

The Tidbinbilla 70-m Radio Telescope is part of the Canberra Deep Space Communication Complex, which is managed by CSIRO Astronomy and Space Science.
The Australia Telescope Compact Array and Parkes radio telescopes are part of the Australia Telescope which is funded by the Commonwealth of Australia for operation as a National Facility managed by CSIRO. This research has made use of: NASA's Astrophysics
Data System Abstract Service; the NASA/
IPAC Infrared Science Archive (which is operated by the Jet Propulsion
Laboratory, California Institute of Technology, under contract with
the National Aeronautics and Space Administration); the SIMBAD data base, operated at CDS, Strasbourg,
France; and data products from the SAGE
survey, which is a legacy science program of the {\em Spitzer Space
  Telescope}, funded by the National Aeronautics and Space
Administration.

\end{document}